\documentclass[10pt]{IEEEtran}

\ifCLASSINFOpdf

\else

\fi

\usepackage{setspace,amsmath,latexsym,cite,amssymb,epsfig,amsfonts}
\usepackage{balance}
\usepackage{enumerate}

\usepackage{algorithm,algpseudocode,float}
\usepackage{lipsum}
\usepackage{xcolor}

\makeatletter
\newenvironment{breakablealgorithm}
  {
   \begin{center}
     \refstepcounter{algorithm}
     \hrule height.8pt depth0pt \kern2pt
     \renewcommand{\caption}[2][\relax]{
       {\raggedright ##2\par}%
       \ifx\relax##1\relax 
         \addcontentsline{loa}{algorithm}{\protect\numberline{\thealgorithm}##2}%
       \else 
         \addcontentsline{loa}{algorithm}{\protect\numberline{\thealgorithm}##1}%
       \fi
       \kern2pt\hrule\kern2pt
     }
  }{
     \kern2pt\hrule\relax
   \end{center}
  }
\makeatother

\begin{document}
\title{Delay Constrained Buffer-Aided Relay Selection \\in the Internet of Things with Decision-Assisted Reinforcement Learning}

\author{Chong Huang, \IEEEmembership{Student Member, IEEE}, Gaojie Chen, \IEEEmembership{Senior Member, IEEE} and Yu Gong 
\thanks{
\noindent This work was supported by EPSRC grant number EP/R006377/1 (“M3NETs”).}
\thanks{Chong Huang and Gaojie Chen are with School of Engineering, Leicester, University, UK, Email: { $\{$ch481, gaojie.chen$\}$@leicester.ac.uk.}}
\thanks{Y. Gong is with Wolfson School of Mechanical, Electrical and Manufacturing Engineering, Loughborough University, UK, Email: { y.gong@lboro.ac.uk.}}
}

\maketitle
\thispagestyle{empty}
\begin{abstract}
This paper investigates the reinforcement learning for the relay selection in the delay-constrained buffer-aided networks. The buffer-aided relay selection significantly improves the outage performance but often at the price of higher latency. On the other hand, modern communication systems such as the Internet of Things often have strict requirement on the latency. It is thus necessary to find relay selection policies to achieve good throughput performance in the buffer-aided relay network while stratifying the delay constraint. With the buffers employed at the relays and delay constraints imposed on the data transmission, obtaining the best relay selection becomes a complicated high-dimensional problem, making it hard for the reinforcement learning to converge. In this paper, we propose the novel {\em decision-assisted} deep reinforcement learning to improve the convergence. This is achieved by exploring the a-priori information from the buffer-aided relay system. The proposed approaches can achieve high throughput subject to delay constraints. Extensive simulation results are provided to verify the proposed algorithms.


\end{abstract}


\section{Introduction}
\IEEEPARstart{W}{ith} the development of 5G communications, the Internet of Things (IoT) is becoming an increasingly growing topic in the area of wireless networks \cite{liu2018novel,7998600,8386655}. IoT applications include massive deployments of wireless devices, requiring high reliability in wireless links \cite{akpakwu2017survey}. The cooperative relay network is known to improve significantly the communication reliability \cite{sheng2011cooperative}, making it an attractive scheme in the IoT. Both amplify-and-forward (AF) and decode-and-forward (DF) relay networks have been developed \cite{li2015opportunistic}.

It is known that the relay selection is an efficient way to harvest the diversity gains. Various relay selection schemes have been proposed. For example, the traditional max-min scheme selects the best relay with the highest signal-to-noise (SNR) links \cite{bletsas2006simple}. In \cite{sun2009cooperative}, a relay selection scheme combined with feedback and adaptive forwarding in cooperative networks was studied. In \cite{hakim2013single}, the bit error rate performance was improved by selecting a single relay node.

The buffer-aided relay selection has attracted much recent attention \cite{zlatanov2011throughput,zlatanov2014buffer}. A number of buffer-aided relay selection schemes have been proposed. In \cite{ikhlef2012max}, the max-max scheme selects the best source-to-relay and best relay-to-destination links for receiving and transmitting at the relay, respectively. In \cite{krikidis2012buffer} and \cite{6817619}, the DF and AF max-link schemes were proposed to select the transmission link with the highest SNR among all source-to-relay and relay-to-destination links, respectively. The max-link can achieve the full diversity order for independent-identical-distributed (i.i.d.) channels when the buffer size is large enough. In \cite{nakai2017generalized}, the buffer-state-based scheme selects the relay based on both the channel status and buffer states, which has better performance in both throughput and delay than the max-link scheme. In \cite{alkhawatrah2019buffer}, a novel prioritization-based buffer-aided link selection was proposed to seamlessly combine the non-orthogonal multiple access (NOMA) and orthogonal multiple access (OMA) transmission. In \cite{6762955}, the max-ratio relay selection was proposed in the cognitive radio network. In \cite{6746659}, the buffer-aided relay selection was applied to improve the physical layer secrecy.

Applying data buffers at the relays often increases the transmission delay which is a key issue in many modern communication networks \cite{zlatanov2014buffer, zhu2017new}. The trade-off between the delay and throughput was studied in the two-hop relay network \cite{qiao2016statistical}. A buffer-aided link selection scheme applying the {\em max-link} and physical layer network coding was proposed to increase the throughput and decrease the average delay in \cite{tian2015buffer}. A delay reduced buffer-aided scheme was proposed by giving higher priority to select the relay-to-destination links than the source-to-relay links \cite{tian2016buffer}. A novel relay selection scheme was proposed to balance the outage probability and average delay by maintaining the buffer length at the appropriate target in \cite{gong2018using}. In \cite{7350164} and \cite{nomikos2018delay}, the delay and diversity aware selection rules were proposed, respectively. In \cite{el2019delay}, a joint physical-layer and upper-layer buffer-aided scheme was proposed to balance the throughput and delay performance in the secure transmission.

None of the above relay selection schemes is optimum to maximize the throughput under the delay constraint. The relay selection can be regarded as the Markov Decision Process (MDP) \cite{Saha14,su2019cooperative}. The reinforcement learning algorithms (e.g., Q-learning and Sarsa), which do not rely on training samples, is of particular interest to solve the MDP problem \cite{zhang2019deep, sutton2011reinforcement}. The transmission efficiency was improved by applying Q-learning for relay selection in DF cooperative networks \cite{Saha14}. In \cite{jadoon2017relay}, the $Q$-learning based relay selection algorithm was proposed to maximize the total throughput of the network. To improve the quality of learning experience from the feature representation, deep neural networks can be applied to improve Q-learning and Sarsa algorithms \cite{mnih2015nature,mnih2016asynchronous}. The deep reinforcement learning is efficient for high-dimensional state and action spaces problems \cite{mnih2015nature}. In \cite{su2019cooperative}, a deep Q-learning based relay selection algorithm was described to improve the outage performance in the multi-relay network. In \cite{Zou20WCNC}, deep reinforcement learning was applied in relay selection for a two-hop relay network to speed up the convergence of training.

The above relay selection schemes based on the reinforcement learning do not consider the data buffers nor the delay constraint. Employing buffers at the relays will significantly increase the dimension of the learning. On the other hand, because the delay is not observed until the packets arrives at the destination, applying delay constraints will also well complicate the learning. Thus directly applying the reinforcement learning may lead to poor convergence. In this paper, the novel decision-assistant deep reinforcement learning is proposed to improve the convergence. This is achieved by exploring the a-priori information from the buffer-aided relay system. The main contributions of this paper are summarized as following:

\begin{itemize}
 \item As far as the authors' aware, this is the first work to apply the deep reinforcement learning in the buffer-aided relay selection.
 \item Two deep reinforcement algorithms, namely the decision-assisted deep $Q$-learning and Sarsa respectively, are proposed for the buffer-aided relay selection subject to instantaneous delay constraints. This is also different from existing buffer-aided schemes which usually consider average packet delay.
 \item The proposed algorithms are well compared and verified with simulations, in which the decision-assisted deep Sarsa learning has the best performance in the relay selection. Particularly for moderate delay constraints, the decision-assisted deep Sarsa can achieve the highest possible throughput in the two-hop relay network,  making it a very attractive scheme in practice.
 \end{itemize}

The rest of the paper is organized as follows: Section \ref{sec:model} describes the system model; Section \ref{sec:formulation} formulates the problem of the optimum relay selection in the delay-constrained buffer-aided relay network; Section \ref{sec:rule} defines the elements of the reinforcement learning for the relay selection; Section \ref{sec:deepRL} describes the deep reinforcement learning for the relay selection; Section \ref{sec:AssiDRL} proposes the decision-assisted deep reinforcement learning explores the a-priori information in the relay; Section \ref{sec:sim} verifies the proposed algorithms with simulation; finally, Section \ref{sec:con} concludes the paper.

\section{System model} \label{sec:model}
The system model of the buffer-aided relay network is shown in Fig. 1, where there are one source node ($S$), one destination node ($D$), a set of $K$ half-duplex DF relays $R_k$ which is equipped with a data buffer of size $L$. We assume that there is no direct link between $S$ and $D$. We also assume that $S$ receives the  the instantaneous channel state information (CSI) and the buffer states, and makes the decision of the relay selection as in  \cite{Zhang20Access}. Moreover, all channels are assumed to experience quasi-static Rayleigh fading with path loss. The channel gain between node $p$ and $q$ is denoted as $|h_{pq}|^2$ which is exponentially distributed with average as E$|h_{pq}|^2 = h_{pq} d_{pq}^{-\alpha}$, where $h_{pq}$ is the fading coefficient which follows the Rayleigh distribution and $d_{pq}^{-\alpha}$ is the path loss, $d_{pq}$ is the corresponding distance between the two nodes and $\alpha$ is the path loss exponent. We assume all channel coefficients are independently fading which remain unchanged during one time slot and vary independently from one time slot to another \cite{tian2016buffer}.

\begin{figure}[t!]
  \centering
  \centerline{\includegraphics[scale=0.6]{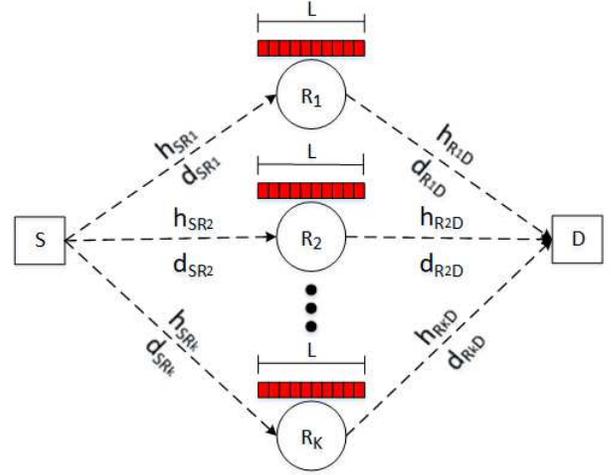}}
 \caption{ System model of the buffer-aided relay network.} \label{fig:systemmodel}
\end{figure}

At one time slot, when a source-to-relay link is selected, a single data packet is transmitted from the source to the corresponding relay $R_k$ and is stored in the buffer. The received signal at $R_k$ is given by
\begin{equation}\label{ysrk}
\begin{split}
   y_{SR_k} = \sqrt{P}h_{SR_k}d_{SR_k}^{-\frac{\alpha}{2}}x_S+n_{R_k},
\end{split}
\end{equation}
where $x_S$ is the data signal from $S$, $n_{R_k}$ is the additive-white-Gaussian-noise (AWGN) noise with variance ${\sigma}_{n}^2$, and $P$ is the transmit power. If a relay-to-destination link is selected, a single data packet from the relay buffer is transmitted to the destination, and the received signal at the destination is given by
\begin{equation}\label{yrkd}
\begin{split}
   y_{R_k{D}} = \sqrt{P}h_{R_k{D}}d_{R_k{D}}^{-\frac{\alpha}{2}}x_{R_k}+n_{D},
\end{split}
\end{equation}
where $x_{R_k}$ is the data signal from $R_k$, $n_{D}$ denotes the AWGN with variance ${\sigma}_{n}^2$ at node $D$. The link capacity between node $p$ and $q$ is given by
\begin{equation}\label{capacityTrans}
\begin{split}
   C_{pq} = {\rm{log_{2}}}\left(1+\frac{P{|h_{pq}|^2}}{{d_{pq}^{\alpha}} {\sigma}_{n}^2}\right).
\end{split}
\end{equation}
The corresponding link is outage if $C_{pq} \leq \eta$, where $\eta$ is the target data rate.

\section{Problem Formulation}\label{sec:formulation}

\begin{figure}[t!]
  \centering
  \centerline{\includegraphics[scale=0.6]{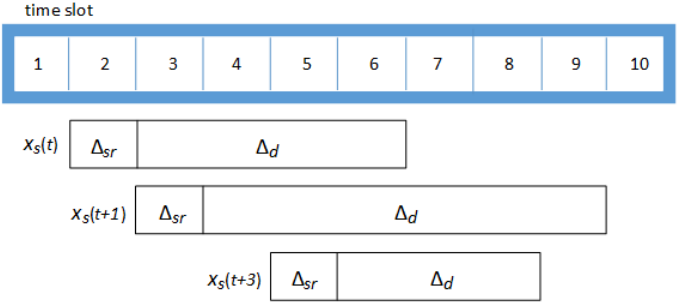}}
 \caption{ An example of delay in the buffer-aided relay network.} \label{fig:delay}
\end{figure}

The delay of a packet is the duration between the packet being transmitted from the source and received at the destination which is given by
\begin{equation}
\begin{aligned}
    \Delta &= \Delta_{sr_k} + \Delta_d= 1 +\Delta_d,
\end{aligned}
\end{equation}
where $\Delta_{sr_k}=1$ which is the transmission time for a successful $S \to R_k$ transmission, and $\Delta_d$ is the delay at $R_k$, which includes the queuing delay and the transmission time for $R_k \to D$. In the two-hop relay network, we have $\Delta \geq 2$. Due to the buffers at the relays, different packet may have different delay and packets may not arrive at the destination with the same order as the transmission order. This is clearly illustrated in Fig. \ref{fig:delay}, where the delays for three packets $x_s(t)$, $x_s(t+1)$ and $x_s(t+3)$ are shown. $x_s(t)$ denotes the packet transmitted from $S$ to $R_k$ at time slot $t$.

For the relay network with $K$ relays, there are $2K$ transmission links. If the buffer at relay $R_k$ is full or empty, the corresponding $S\to R_k$ or $R_k \to D$ is not available for data transmission, respectively. At time slot $t$, we denote $m_{k,0}(t)$ and $m_{k,1}(t)$ as the decision variables for the $S \to R_k$ and $R_k \to D$ transmission respectively, where $m_{k,j}(t) \in \{0,1 \}$. When $m_{k,j}(t)=1$, the corresponding link is selected at time slot $t$, otherwise it is not. At the time slot $t$, for all $k$ and $j$, either only one $m_{k,j}(t)$ is $1$, or all of $m_{k,j}(t)$ are $0$ which corresponds to the outage event that no link can be selected. Therefore we have
\begin{equation}\label{eq:qkide}
    \sum_{k=1}^K\sum_{j=0}^1 m_{k,j}(t)= 0 ~{\rm or}~ 1.
\end{equation}

At time slot $t$, if $R_k \to D$ is selected (i.e. $m_{k,1}(t)=1$), one packet of throughput is observed at the destination. The optimum relay selection policy satisfies
\begin{equation} \label{eq:cost}
    \begin{aligned}
        & \max \sum_{t=1}^N  \sum_{k=1}^K  m_{k,1}(t) \cdot \mu(C_{R_k,D}(t) > \eta), \\
        & \quad {\rm s.t.} \quad \Delta(t) \leq \Delta_{o}, \\
        & \qquad \quad m_{k,j}(t) = 0 ~{\rm or}~ 1, \quad t=1, \cdots, N\\
        & \qquad \quad \sum_{k=1}^K\sum_{j=0}^1 m_{k,j}(t) = 0 ~{\rm or}~ 1, \quad t=1, \cdots, N
    \end{aligned}
\end{equation}
where $\mu(.)=1$ if the enclosed holds and $0$ if otherwise, $N$ is the number of the time slots observed at the destination, $\Delta(t)$ is the delay for the packet arriving at $D$ at time slot $t$, and $\Delta_o$ is the target packet delay. The constraint $\Delta(t) \leq \Delta_{o}$ ensures that only the receiving packets with delay smaller than $\Delta_o$ contribute to the overall throughput.

The solution of \eqref{eq:cost} depends on various factors including instantaneous channel gains, buffer states and experienced delays for every packet in the buffers. These factors are time varying and often have conflicting requirements in the relay selection. For example, selecting the strongest link may not satisfy the delay constraint, or it may cause the buffer overflow or empty, which again leads to less available links for selection. Moreover, the system is time-varying which makes the problem be more complicated. Solving in \eqref{eq:cost} is in general a complicated task, if possible, particularly when the buffer size or relay number is large. This motives us to investigate machine learning solutions in the following Sections.

\begin{figure}[t!]
  \centering
  \centerline{\includegraphics[scale=0.6]{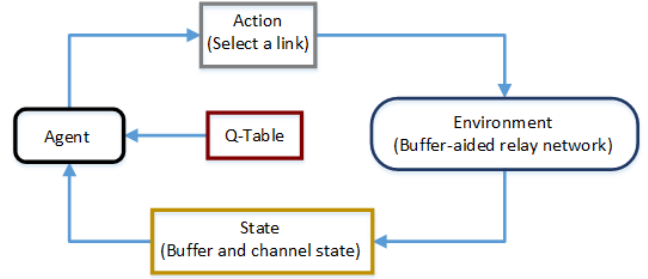}}
 \caption{ The framework of the reinforcement learning.} \label{fig:DRL}
\end{figure}

Because there is no analytical solution to \eqref{eq:cost}, the labelled data cannot be obtained to apply the training-based machine learning. On the other hand, the buffer-aided relay selection problem can be modelled as an MDP \cite{sun2009cooperative}, and the reinforcement learning can be applied. The framework of the reinforcement learning in buffer-aided relay networks is shown in Fig. \ref{fig:DRL}. In the next section, we will define the elements of the reinforcement learning for the buffer-aided relay selection, based on which the deep reinforcement is introduced in Section \ref{sec:deepRL}. And in Section \ref{sec:AssiDRL} the decision-assisted deep reinforcement learning is proposed.

\section{Elements of the Reinforcement learning for the buffer-aided relay selection}\label{sec:rule}

The elements of the reinforcement learning consist of environment, state, action, rewards and agent. They must be carefully defined to efficiently reveal the buffer-aided relay selection system to ensure the implementation of the reinforcement learning.

\subsection{Environment and action} \label{sec:action}

The {\em environment} is the buffer-aided relay selection network. The {\em action} is to select a link for data transmission, which is equivalent to determining $m_{k,j}$ in \eqref{eq:qkide}. For the relay network with $K$ relays, there are $2K$ transmission links. At any time slot, either one of the links is selected for transmission which corresponds to $2K$ possible actions, or no link can be selected which corresponds to another action, making the total number of actions to $2K+1$.

\subsection{State}\label{sec:state}

The relay selection depends on the instantaneous CSI and buffer states which shall both be included in the {\em state} for the reinforcement learning. The buffer state at time slot $t$ is represented as $ \{l_{t,1},. . . , l_{t,K}\}$, where $l_{t,k}$ is the buffer length (i.e. the number of packet in the buffer) for the $k$th buffer at time slot $t$. When a buffer is full (i.e $l_{t,k}=L$) or empty ($l_{t,k}=0$), the corresponding $S\to R_k$ or $R_k\to D$ link is not available for data transmission, respectively. On the other hand, because the CSI-s take continuous values, directly applying the CSI-s will cause infinite number of {\em states}, which makes it very hard for the reinforcement learning to converge.

{\em Remark 1: At time slot $t$, a link is invalid for selection if:
\begin{itemize}
    \item The corresponding link capacity cannot support the target data rate $\eta$, i.e. $C_{pq}(t) \leq \eta$.
    \item Or for a $S\to R_k$ link, the buffer is full, i.e. $l_{t,k} = L$.
    \item Or for a $R_k\to D$ link, the buffer is empty, i.e. $l_{t,k} = 0$.
\end{itemize}
If an action is to select an invalid link, the action is also deemed as invalid.}

Therefore, at time slot $t$ and for relay $R_k$, we use $c_{t,k}$ to specify the validness of the corresponding links as
\begin{itemize}
    \item $c_{t,k}=1$: only the $S\to R_k$ link is valid for selection;
    \item $c_{t,k}=2$: only the $R_k \to D$ link is valid;
    \item $c_{t,k}=3$: both links are valid;
    \item $c_{t,k}=4$: none of the two links is valid.
\end{itemize}

The {\em state} in the environment at time slot $t$ is then defined as
\begin{equation}\label{state2}
\begin{split}
   s_{t} = \{l_{t,1},\cdots , l_{t,K},c_{t,1}, \cdots , c_{t,K}\}.
\end{split}
\end{equation}
With the buffer size $L$ and relay number $K$, the total number of states is $(4(L+1))^K$.

\subsection{Rewards} \label{sec:Qtable}
The purpose of the learning is to maximize the delay constrained throughput as is shown in \eqref{eq:cost}. At one time slot when a link is selected (corresponding to an {\em action} taken), if a packet is successfully transmitted to the destination $D$ within the target delay $\Delta_o$, a reward is given to the corresponding action. 
We note that selecting a $R_k\to D$ link may not necessarily lead to an award because either the link may be in outage or the transmitted packet has larger delay than $\Delta_o$.

\begin{figure}[t!]
  \centering
  \centerline{\includegraphics[scale=0.6]{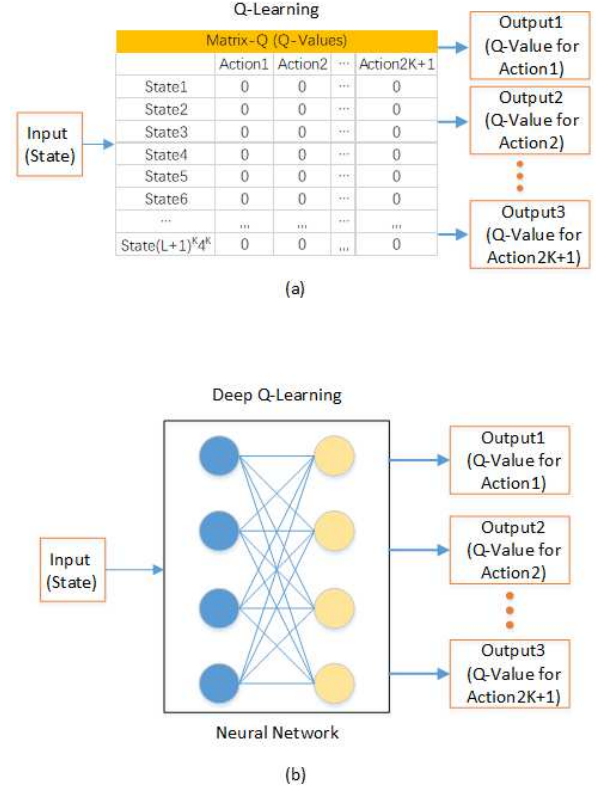}}
 \caption{ $Q$-table {\em vs} deep neural network.} \label{fig:compare}
\end{figure}

The $Q$-table is used to store the accumulated rewards (namely the $Q$-values) for every action at all states. In the buffer-aided relay network, because there are $(4(L+1))^K$ states and $2K+1$ actions, the $Q$-table is a $(4(L+1))^K$ by $2K+1$ matrix, as is illustrated in Fig. \ref{fig:compare} (a).

Because the size of the $Q$-table can be very large when the buffer size $L$ or relay number $K$ is large, directly updating the $Q$-table may result in severe overfitting and slow convergence. In the deep reinforcement learning, the deep neural network is used to realize the $Q$-table as is shown in Fig. \ref{fig:compare} (b), where the input layer is the state $s_t$, and the output layer generates ($2K+1$) $Q$-values corresponding to the ($2K+1$) possible actions at state $s_t$. To be specific, if we use the Q-table in reinforcement learning, the algorithm will find the maximum value in the Q-table for a given state, and take the corresponding action from the Q-table. On the other hand, if we apply the deep neural network, the algorithm will input the state into the deep neural network and get the evaluation values for each action-state pair, and then take the action with the maximum value from the neural network. The advantage of deep neural network is to avoid building a huge Q-table if the number of state is quite large.


\subsection{Agent} \label{sec:agent}
The agent generates the `experiences' of data for the $Q$-learning by interacting with the environment. This paper investigates the $Q$-learning and Sarsa \cite{mnih2016asynchronous}. The experiences for each learning method are generated as following.
\\

\noindent{$\bullet~$\em \textbf{Generate Q-learning experiences }} \label{sec:QTLea}

\begin{enumerate}
    \item Suppose at time slot $t$, the state is $s_t$. From the current $Q$-table denoted as $Q_t(.)$ (i.e. the prediction network as is shown in the next section), the agent decides the next action: i.e.  which link (or whether there is a link) is selected for data transmission. The $\varepsilon$-greedy strategy is often applied to determine the action $a_t$ at state $s_t$:
    \begin{equation}\label{eq:Qact}
        a_t = \left\{
        \begin{array}{ll}
             \arg \max_a Q_t(s_t, a), & {\rm with ~ prob.~} (1-\varepsilon)  \\
             random~selection, & {\rm otherwise}
        \end{array},
        \right.
    \end{equation}
    where $0\leq \varepsilon \leq 1$. In this paper, $\varepsilon$ is set as following
    \begin{equation}\label{greedy}
        \varepsilon = {\rm max} (f^{N_{ite}-1}, \varepsilon_{min}),
    \end{equation}
    where $f \in (0, 1)$ which is the decay factor , $N_{ite}$ the number of training iterations, and $\varepsilon_{min} \in (0, 1)$ which is the minimum value for $\varepsilon$. From \eqref{greedy}, $\varepsilon$ is initially set to $1$ for good exploration, and gradually increased with the learning iterations.
\item Once the action $a_t$ is chosen, the reward $r_{s_t, a_t}$ is given. If $a_t$ leads to a $S\to R_k$ or a $R_k\to D$ link selection, the corresponding buffer length is increased or decreased by one, respectively, and otherwise the buffer length remains unchanged. On the other hand, the channel states vary independently from one time slot to another. Then the state transits to $s_{t+1}$ based on the new buffer-lengths and channel states.
\item One experience is then generated as
\begin{equation}\label{eq:traData}
\{ s_t, a_t, r_{s_t, a_t}, s_{t+1} \}.
\end{equation}
\item Go back to step 1) to repeat the process with state $s_{t+1}$, and generate another experience. \\
\end{enumerate}


\noindent{$\bullet~$\em \textbf{Generate Sarsa experiences}}

In the Sarsa, the action for the current state is chosen based on the prediction at the previous time slot. Because the Sarsa predicts the action in one time slot ahead, it sticks more to the selected action. Thus the Sarsa has more {\em exploitation} but less {\em exploration} than the $Q$-learning.

The experiences in Sarsa are generated as following:

\begin{enumerate}
\item Suppose at the time slot $t$, the state is $s_t$. Unlike the $Q$-learning, the action $a_t$, which is predicted at the previous time slot $t-1$, is applied for $s_t$.
\item For the action $a_t$, the reward $r_{s_t, a_t}$ is given, and the state transits to $s_{t+1}$.

    \item Using the current $Q$-table, $Q_t$, to predict the action for $s_{t+1}$ for the next time slot $t+1$ as
    \begin{equation}\label{eq:Saa1}
            a_{t+1} = \left\{
            \begin{array}{ll}
                 \arg \max_a Q_{t}(s_{t+1}, a), & {\rm with ~ prob.~} (1-\varepsilon)  \\
                 random~selection, & {\rm otherwise}
            \end{array}.
            \right.
    \end{equation}

\item One experience is generated as
\begin{equation}\label{eq:traSasa}
\{ s_t, a_t, r_{s_t, a_t}, s_{t+1}, a_{t+1} \}.
\end{equation}
Unlike the $Q$-learning, the predicted action $a_{t+1}$ is also included in the experience.


\item Go back to step 1) to repeat the process with state $s_{t+1}$, and generate another experience.
\end{enumerate}

\section{Deep reinforcement learning for the buffer-aided relay selection} \label{sec:deepRL}

\begin{figure}[t!]
  \centering
  \centerline{\includegraphics[scale=0.6]{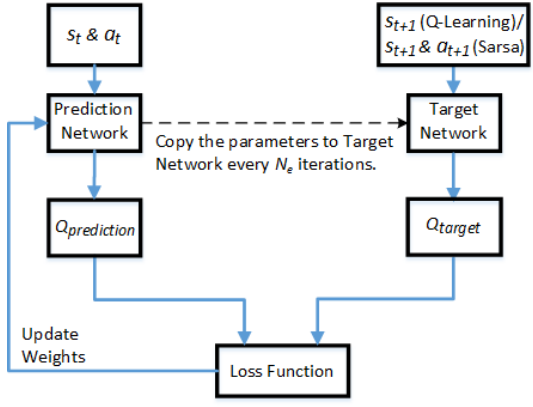}}
 \caption{ The system model of the deep reinforcement learning.} \label{fig:nn}
\end{figure}

The system model of the deep reinforcement learning is shown in Fig. \ref{fig:nn}, which consists of the prediction and target deep neural networks, generating the $Q$-values for the current and next time states $s_t$ and $s_{t+1}$, respectively. The prediction and target networks are updated following the below three steps, where either $Q$-learning or Sarsa can be used.\\

\noindent {\em Step 1 - Generate experiences.}

The agent applies the relay selection in the buffer-aided relay system for $N_g$ time slots, and generate $N_g$ experiences for either $Q$-learning or Sara.

In Step 1, the prediction network coefficients remain fixed, while the target network is not involved. \\

\noindent {\em Step 2 - Update the prediction network.}

\begin{figure*}[t!]
  \centering
  \centerline{\includegraphics[scale=0.6]{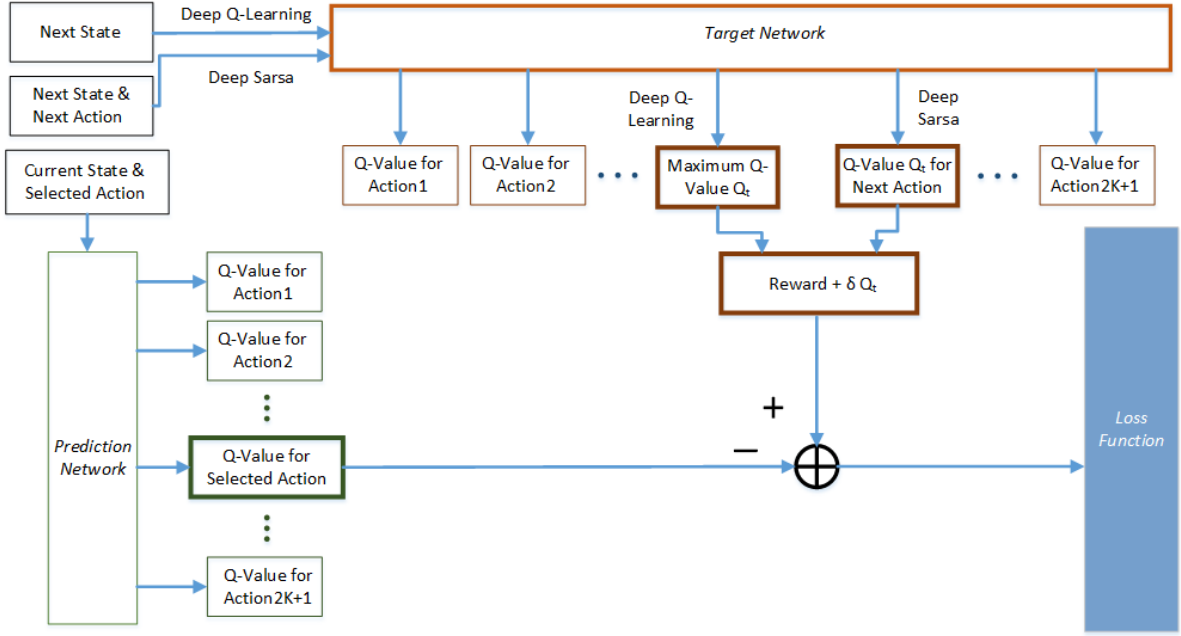}}
 \caption{ Update the prediction network with $Q$-learning or Sarsa} \label{fig:WD}
\end{figure*}

From the $N_g$ experiences generated in Step 1, $N_p$ experiences are randomly chosen \cite{mnih2013DQN} to update the prediction network as is shown in Fig. \ref{fig:WD}. The target network remains unchanged in Step 2.

\begin{itemize}

    \item In the predication network, for the $i$-th experience, the output is given by
    \begin{equation}\label{eq:prei}
        Pre(i) = Q_{prediction}(s_t^{(i)}, a_t^{(i)}),
    \end{equation}
    which is the $Q$-value for taking action $a_t^{(i)}$ at $s_t^{(i)}$ based on the prediction network.

    \item In the target network, for the $i$-th experience, if the $Q$-learning is applied, the output is
    \begin{equation}\label{eq:tarQ}
        Tar(i) = \max_a~ Q_{target}(s_{t+1}^{(i)}, a),
    \end{equation}
    which is the largest $Q$-value with respect to all actions of the target network at $s_{t+1}^{(i)}$.

    On the other hand, if the Sarsa is applied, and the output of the target network is
    \begin{equation}\label{eq:tarS}
        Tar(i) = Q_{target}(s_{t+1}^{(i)}, a_{t+1}^{(i)}),
    \end{equation}
    which is the $Q$-value for the specified action $a_{t+1}^{(i)}$.

    \item Repeat the above procedure for all of the selected $N_p$ experiences and form the cost function as
    \begin{equation}\label{eq:QcosPre}
    \begin{aligned}
        L_{Q} =  \sum_{i=1}^{N_p} \left ( \delta \cdot Tar(i)
         + r_{s_t, a_t}^{(i)} - Pre(i) \right)^2,
    \end{aligned}
    \end{equation}
    where is $\delta$ is the discount factor.

    \item Based on \eqref{eq:QcosPre}, the coefficients of the prediction networks are updated once with the gradient descent search algorithm \cite{mnih2016asynchronous,lange2012batch}. For better convergence, the Adam algorithm is used to realize adaptive learning rate in this paper \cite{kingma2014adam,goodfellow2016deep}. \\
\end{itemize}

\noindent {\em Step 3 - Update the target network}

After repeating {\em Step 1} and {\em 2} for $N_e$ iterations, copy the prediction network coefficients to the target network. Go back to {\em Step 1} for another round until the $Q$-table (i.e. the networks) converges or the maximum number of rounds is reached.

\section{Decision-assisted deep reinforcement learning} \label{sec:AssiDRL}

As is shown in {\em Remark 1} in Section \ref{sec:state}, at the state $s_t$, not all of the  ($2K+1$) actions are valid for selection.
The straightforward way to handle the invalid actions is to ignore them. To be specific, at the stage of generating experiences, if an invalid action is selected for a given state, either the corresponding experience is simply ignored, or the valid action with the highest Q-value is used in the experience (even though an invalid action has higher Q-value at the state).

While this provides a simple way to avoid the invalid actions, the learning performance is not promising. Because in the deep reinforcement learning, the Q-values for all actions (both valid and invalid) are from the same prediction neural network, updating the neural network weights for one action will also affect the Q-values for the other actions. This is different from direct updating the Q-table where the Q-values for different actions are updated independently. Furthermore, an invalid action in one state may become valid in another state (and vice versa). Therefore, by simple ignoring the invalid actions in the learning, the neural network weights are only updated for valid actions. The Q-values for the invalid actions are not directly updated and so may not converge to the desirable small values. This results in slow convergence or converging to local minimums, if it converges at all.

In this section, we investigate two methods to explore the a-priori information about the invalid actions to improve the reinforcement learning.

\subsection{Punishment with negative rewards}

In Section \ref{sec:Qtable}, a reward is only given to an action if it leads to one packet arriving at the destination within the target delay. All other actions, whether they are valid or not, will not be awarded without any difference. This will take a long time for the learning to avoid selecting invalid links, leading to slow convergence (if it converges at all).

To overcome this problem, one common way in the reinforcement learning is to introduce the `negative reward'. In the buffer-aided relay selection, this is to apply negative rewards to `punish' the invalid actions if they are selected. To achieve this, we divide all of the ($2K+1$) possible actions at state $s_t$ into three categories:
\begin{itemize}
    \item ${\mathcal VR}_{s_t}$: the valid actions at $s_t$ which leads to one packet successfully arriving at the destination from a relay within the target delay time.
    \item ${\mathcal VZ}_{s_t}$: all other valid actions without leading to packet arriving at the destination within the target delay. These include valid actions to choose the $S\to R_k$ links or  the $R_k \to D$ links with higher delay than the target delay.
    \item ${\mathcal {\bar V}}_{s_t}$: all invalid actions at state $s_t$, as is specified in {\em Remark 1} in Section \ref{sec:state}.

\end{itemize}

Therefore at state $s_t$, if the action $a_t$ is taken, the reward is given by
\begin{equation}\label{eq:rstpu}
\begin{aligned}
    r_{s_t, a_t} : \left \{
    \begin{array}{ll}
         > 0,  & a_t \in {\mathcal VR}_{s_t} \\
         = 0, &  a_t \in {\mathcal VZ}_{s_t}\\
         < 0, &  a_t \in {\mathcal {\bar V}}_{s_t}
    \end{array}
    \right.
\end{aligned}
\end{equation}

Applying punishments with negative rewards as in \eqref{eq:rstpu} makes the learning quickly avoid the invalid actions. However, the punishment can be too harsh so that the learning may regard actions without punishments as good enough decisions. Particularly, because the neural networks weights shall be randomly initialized \cite{goodfellow2016deep}, the networks may not generate high enough rewards for `optimal' actions after many iterations. This makes it vulnerable for the learning to converge to local optimums. This will be well verified in the simulations later. A better way to handle the invalid actions is needed.

\subsection{Decision assisted learning} \label{sec:DAL}

In the deep reinforcement learning, the $i$th experience for the $Q$-learning and Sarsa are given by $\{ s_t^{(i)}, a_t^{(i)}, r_{s_t, a_t}^{(i)}, s_{t+1}^{(i)} \}$ and $\{ s_t^{(i)}, a_t^{(i)}, r_{s_t, a_t}^{(i)}, s_{t+1}^{(i)}, a_{t+1}^{(i)} \}$, respectively. For the $i$th experience, the prediction and target networks generate $Pre(i)$ and $Tar(i)$, as is shown in \eqref{eq:prei} and \eqref{eq:tarQ} (or \eqref{eq:tarS} for Sarsa), respectively. This is equivalent to one pair of labelled training data as
\begin{equation}\label{eq:0Tpr}
   \{Pre(i), ~~\delta \cdot Tar(i) + r_{s_t, a_t} ^{(i)}\}.
\end{equation}
Because the target network itself is estimated, $Tar(i)$ is also an estimation but not the true target $Q$-value. This is the reason why the deep reinforcement learning always converges well slower than its traditional training-based counterpart.

On the other hand, for the invalid actions (i.e. $a \in {\mathcal {\bar V}}_{s_t^{(i)}}$ as is shown in {\em Remark 1}), the true target $Q$-values shall be zero, because the invalid actions do not lead to any rewards either at the current time or in the future. With these consideration, besides \eqref{eq:0Tpr}, we can formulate extra `training pairs' for one experience as
\begin{equation}\label{eq:tpari}
\begin{aligned}
  \{ Q_{prediction}(s_t^{(i)}, a), ~~0 \},
  \qquad a \in {\mathcal {\bar V}}_{s_t^{(i)}}
\end{aligned}
\end{equation}
where $ Q_{prediction}(s_t^{(i)}, a)$ is defined in \eqref{eq:prei} which is the $Q$-value output of the prediction network for action $a$ at state $s_t^{(i)}$.

\begin{figure*}[t!]
  \centering
  \centerline{\includegraphics[scale=0.6]{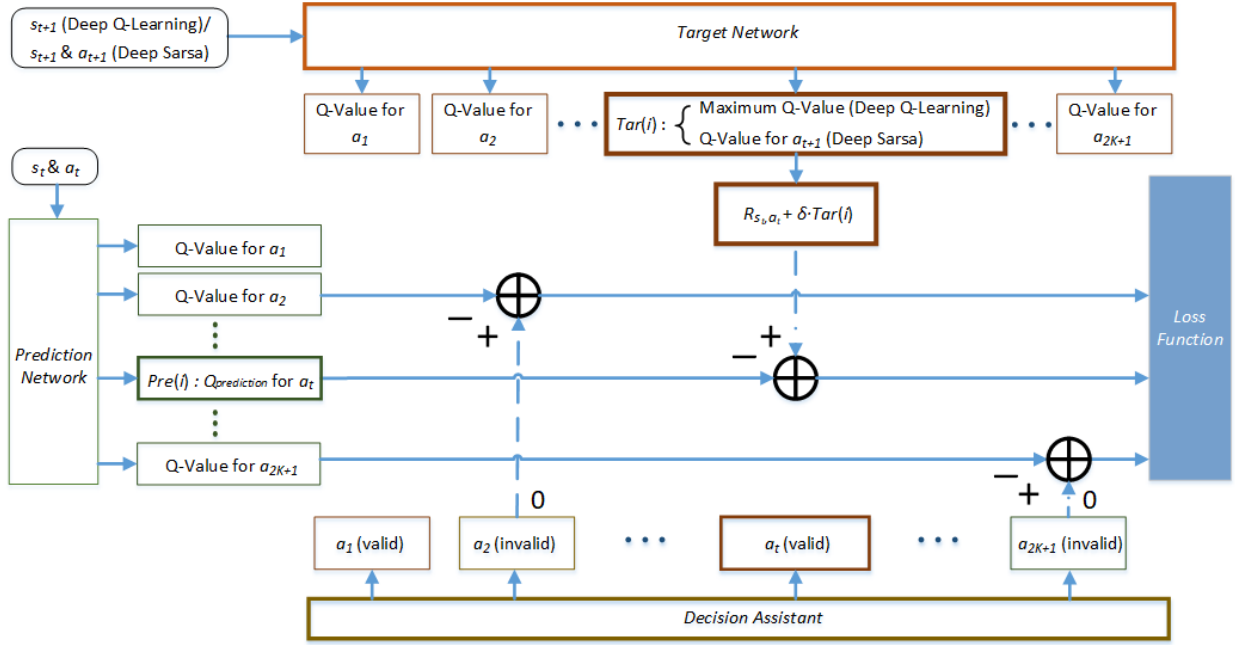}}
 \caption{ Training deep neural network with Decision Assistant, where 0 from Decision Assistant denotes the corresponding link cannot be selected, and 1 denotes the corresponding link can be selected or no link can be selected} \label{fig:WDA}
\end{figure*}

As is illustrated in Fig. \ref{fig:WDA}, we include all training pairs in \eqref{eq:0Tpr} and \eqref{eq:tpari} for all of the $N_p$ experiences to form the cost function as
\begin{equation}\label{eq:DiCost}
    \begin{aligned}
        L_{D} =  \sum_{i=1}^{N_p} & \left\{ \left ( \delta \cdot Tar(i)
         + r_{s_t, a_t}^{(i)} - Pre(i) \right)^2 \right. \\
         & + \sum_{a \in {\mathcal {\bar A}}_{s_t^{(i)}}} \left.  \left ( Q_{prediction}(s_t^{(i)}, a) \right )^2 \right \}.
    \end{aligned}
\end{equation}
The cost function in \eqref{eq:DiCost} is then used to update the prediction network, where either the $Q$-learning or Sarsa can be applied. This is called the decision-assisted deep reinforcement learning in this paper, as the learning is assisted by exploring the a-priori information about invalid actions.

Fig. \ref{fig:WD} and \ref{fig:WDA} clearly show the significant difference between with and without the decision-assisted deep reinforcement learning. In the former case, the prediction network weights are updated only based on the selected actions in every experience. On the contrary, in the latter case, the zero target $Q$-values for the invalid actions are always applied no matter whether they are selected or not in all experiences. Therefore, the neural network weights are constantly `trained' to output zero Q-values for the invalid actions for every experience. As a result, the reward-based `reinforcement learning' only needs to explore in the valid actions. This well reduces the exploration dimension for the learning, leading to faster convergence to the desired policies. This is of particular interest to the Sarsa, because the Sarsa {\em exploits} more but {\em explores} less than its $Q$-learning counterpart. This will be well verified in the simulations later.

\subsection{The algorithm summary} \label{sec:algorithm}

The {\em decision-assist} can be applied with both deep $Q$-learning and Sarsa algorithm, leading to the following two algorithms:
\begin{itemize}
    \item {\em DAD-QL}: The decision-assisted deep $Q$-learning algorithm.
    \item {\em DAD-Sarsa}:  The decision-assisted deep Sarsa learning algorithm.
\end{itemize}

The {\em  DAD-QL} and {\em DAD-Sarsa} algorithms are summarized as following.
\begin{breakablealgorithm}
\caption{\textbf{DAD-QL Algorithm}: Decision-assisted deep Q-Leaning for the delay constrained buffer-aided relay selection}\label{alg:MyAlgorithm1}
\begin{enumerate}[\indent 1.]
\item Initialize the environmental variables.
\item Repeat:
\item \hphantom{0}\hphantom{0}{\bf For} $v = 1, \cdots, N_e$ do:\\
\item \hphantom{0}\hphantom{0}\hphantom{0}{\bf (Step 1: Generate {\em training} experiences):}
\item \hphantom{0}\hphantom{0}\hphantom{0}\hphantom{0}{\bf For} $t = 1,  \cdots, N_g$ do:
\item \hphantom{0}\hphantom{0}\hphantom{0}\hphantom{0}\hphantom{0}\hphantom{0}For state $s_t$, and based on the prediction network,
\item[]\hphantom{0}\hphantom{0}\hphantom{0}\hphantom{0}\hphantom{0}\hphantom{0}\hphantom{0}use the $\varepsilon$-greedy strategy \eqref{eq:Qact} to decide $a_{t}$.
\item \hphantom{0}\hphantom{0}\hphantom{0}\hphantom{0}\hphantom{0}\hphantom{0}Obtain the next state $s_{t+1}$ and the reward $r_{s_t,a_t}$.
\item \hphantom{0}\hphantom{0}\hphantom{0}\hphantom{0}\hphantom{0}\hphantom{0}Generate one experience of {\em training} as
\item[]\hphantom{0}\hphantom{0}\hphantom{0}\hphantom{0}\hphantom{0}\hphantom{0}\hphantom{0}\hphantom{0} \hphantom{0}\hphantom{0} $\{ s_t, a_t, r_{s_t, a_t}, s_{t+1} \}$.
\item \hphantom{0}\hphantom{0}\hphantom{0}\hphantom{0}{\bf end} of ``for $t = 1,  \cdots, N_g$ do"\\

\item \hphantom{0}\hphantom{0}\hphantom{0}{\bf (Step 2: Update the prediction network):}
\item \hphantom{0}\hphantom{0}\hphantom{0}\hphantom{0}Randomly choose $N_p$ experiences from Step 1.
\item \hphantom{0}\hphantom{0}\hphantom{0}\hphantom{0}{\bf For} $i=1, \cdots, N_p$ do:
\item \hphantom{0}\hphantom{0}\hphantom{0}\hphantom{0}\hphantom{0}\hphantom{0}Take the $i$th experience: $\{ s_t^{(i)}, a_t^{(i)}, r_{s_t, a_t}^{(i)}, s_{t+1}^{(i)} \}$.
\item \hphantom{0}\hphantom{0}\hphantom{0}\hphantom{0}\hphantom{0}\hphantom{0}Obtain the prediction network output as:
\item[]\hphantom{0}\hphantom{0}\hphantom{0}\hphantom{0}\hphantom{0}\hphantom{0}\hphantom{0}\hphantom{0}\hphantom{0}\hphantom{0}\hphantom{0}$Pre(i) = Q_{prediction}(s_t^{(i)}, a_t^{(i)})$.
\item \hphantom{0}\hphantom{0}\hphantom{0}\hphantom{0}\hphantom{0}\hphantom{0}Obtain the target network output as:
\item[]\hphantom{0}\hphantom{0}\hphantom{0}\hphantom{0}\hphantom{0}\hphantom{0}\hphantom{0}\hphantom{0}\hphantom{0}\hphantom{0}\hphantom{0}$ Tar(i) = \max_a~ Q_{target}(s_{t+1}^{(i)}, a)$.
\item \hphantom{0}\hphantom{0}\hphantom{0}\hphantom{0}\hphantom{0}\hphantom{0}Form decision-assisted {\em training} pairs as:
\begin{equation*}
\begin{aligned}
  ~~~~~~~~~&\{ Pre(i), ~\delta \cdot Tar(i) + r_{s_t, a_t}^{(i)} \}, \\
    &\{ Q_{prediction}(s_t^{(i)}, a), ~0 \}, \quad a \in {\mathcal {\bar A}}_{s_t^{(i)}}.
\end{aligned}
\end{equation*}
\item \hphantom{0}\hphantom{0}\hphantom{0}\hphantom{0}{\bf end} of ``For $i=1, \cdots, N_p$ do"
\item \hphantom{0}\hphantom{0}\hphantom{0}\hphantom{0}Form the cost function as:
\begin{equation*}
    \begin{aligned}
        ~~~~~~~~~L_{D} =  \sum_{i=1}^{N_p} & \left\{ \left ( \delta \cdot Tar(i)
         + r_{s_t, a_t}^{(i)} - Pre(i) \right)^2 \right. \\
         ~~~~~~~~~& + \sum_{a \in {\mathcal {\bar A}}_{s_t^{(i)}}} \left.  \left ( Q_{prediction}(s_t^{(i)}, a) \right )^2 \right \}.
    \end{aligned}
\end{equation*}
\item \hphantom{0}\hphantom{0}\hphantom{0}\hphantom{0}Update the prediction network based on $L_D$.

\item \hphantom{0}\hphantom{0}\textbf{end} of ``For $v = 1, \cdots, N_e$ do:" \\

\item \hphantom{0}{\bf (Step 3: Update the target network):}
\item \hphantom{0}\hphantom{0}Copy the prediction network coefficients to the
\item[]\hphantom{0}\hphantom{0}\hphantom{0}target network.
\item Until the end of learning.
\end{enumerate}
\end{breakablealgorithm}

\begin{breakablealgorithm}
\caption{\textbf{DAD-Sarsa Algorithm}: Decision-assisted deep Sara learning for the delay constrained buffer-aided relay selection}\label{alg:MyAlgorithm2}
\begin{enumerate}[\indent 1.]
\item Initialize the environmental variables.
\item Repeat:
\item \hphantom{0}\hphantom{0}{\bf For} $v = 1, \cdots, N_e$ do:\\
\item \hphantom{0}\hphantom{0}\hphantom{0}{\bf (Step 1: Generate {\em training} experiences):}
\item \hphantom{0}\hphantom{0}\hphantom{0}\hphantom{0}{\bf For} $t = 1,  \cdots, N_g$ do:
\item \hphantom{0}\hphantom{0}\hphantom{0}\hphantom{0}\hphantom{0}\hphantom{0}At $s_t$, take the action $a_t$ predicted at ($t-1$).
\item \hphantom{0}\hphantom{0}\hphantom{0}\hphantom{0}\hphantom{0}\hphantom{0}Obtain state $s_{t+1}$ and reward $r_{s_t,a_t}$.
\item \hphantom{0}\hphantom{0}\hphantom{0}\hphantom{0}\hphantom{0}\hphantom{0}For $s_{t+1}$, and based on the prediction network,
\item \hphantom{0}\hphantom{0}\hphantom{0}\hphantom{0}\hphantom{0}\hphantom{0}use the $\varepsilon$-greedy strategy \eqref{eq:Saa1} to predict the next
\item[]\hphantom{0}\hphantom{0}\hphantom{0}\hphantom{0}\hphantom{0}\hphantom{0} action $a_{t+1}$.
\item \hphantom{0}\hphantom{0}\hphantom{0}\hphantom{0}\hphantom{0}\hphantom{0}Generate one experience of {\em training} as
\item[]\hphantom{0}\hphantom{0}\hphantom{0}\hphantom{0}\hphantom{0}\hphantom{0}\hphantom{0}\hphantom{0}\hphantom{0}\hphantom{0}\hphantom{0}\hphantom{0}$\{ s_t, a_t, r_{s_t, a_t}, s_{t+1}, a_{t+1} \}$.
\item \hphantom{0}\hphantom{0}\hphantom{0}\hphantom{0}{\bf end} of ``for $t = 1,  \cdots, N_g$ do"\\
\item \hphantom{0}\hphantom{0}\hphantom{0}{\bf (Step 2: Update the prediction network):}
\item \hphantom{0}\hphantom{0}\hphantom{0}\hphantom{0}Randomly choose $N_p$ experiences from Step 1.
\item \hphantom{0}\hphantom{0}\hphantom{0}\hphantom{0}{\bf For} $i=1, \cdots, N_p$ do:
\item \hphantom{0}\hphantom{0}\hphantom{0}\hphantom{0}\hphantom{0}\hphantom{0}Take the $i$th experience:
\item[]\hphantom{0}\hphantom{0}\hphantom{0}\hphantom{0}\hphantom{0}\hphantom{0}\hphantom{0}\hphantom{0}\hphantom{0}\hphantom{0}\hphantom{0}$\{ s_t^{(i)}, a_t^{(i)}, r_{s_t, a_t}^{(i)}, s_{t+1}^{(i)}, a_{t+1}^{(i)} \}$.
\item \hphantom{0}\hphantom{0}\hphantom{0}\hphantom{0}\hphantom{0}\hphantom{0}Obtain the prediction network output as:
\item[]\hphantom{0}\hphantom{0}\hphantom{0}\hphantom{0}\hphantom{0}\hphantom{0}\hphantom{0}\hphantom{0}\hphantom{0}\hphantom{0}\hphantom{0}$Pre(i) = Q_{prediction}(s_t^{(i)}, a_t^{(i)})$.
\item \hphantom{0}\hphantom{0}\hphantom{0}\hphantom{0}\hphantom{0}\hphantom{0}Obtain the target network output as:
\item[]\hphantom{0}\hphantom{0}\hphantom{0}\hphantom{0}\hphantom{0}\hphantom{0}\hphantom{0}\hphantom{0}\hphantom{0}\hphantom{0}\hphantom{0}$Tar(i) = Q_{target}(s_{t+1}^{(i)}, a_{t+1}^{(i)})$.
\item \hphantom{0}\hphantom{0}\hphantom{0}\hphantom{0}\hphantom{0}\hphantom{0}Form decision-assisted {\em training} pairs as:
\begin{equation*}
\begin{aligned}
  ~~~~~~~~~&\{ Pre(i), ~\delta \cdot Tar(i) + r_{s_t, a_t}^{(i)} \}, \\
    &\{ Q_{prediction}(s_t^{(i)}, a), ~0 \}, \quad a \in {\mathcal {\bar A}}_{s_t^{(i)}}.
\end{aligned}
\end{equation*}
\item \hphantom{0}\hphantom{0}\hphantom{0}\hphantom{0}{\bf end} of ``For $i=1, \cdots, N_p$ do"
\item \hphantom{0}\hphantom{0}\hphantom{0}\hphantom{0}Form the cost function as:
\begin{equation*}
    \begin{aligned}
        ~~~~~~~~~L_{D} =  \sum_{i=1}^{N_p} & \left\{ \left ( \delta \cdot Tar(i)
         + r_{s_t, a_t}^{(i)} - Pre(i) \right)^2 \right. \\
         ~~~~~~~~~& + \sum_{a \in {\mathcal {\bar A}}_{s_t^{(i)}}} \left.  \left ( Q_{prediction}(s_t^{(i)}, a) \right )^2 \right \}.
    \end{aligned}
\end{equation*}
\item \hphantom{0}\hphantom{0}\hphantom{0}\hphantom{0}Update the prediction network based on $L_D$.

\item \hphantom{0}\hphantom{0}\textbf{end} of ``For $v = 1, \cdots, N_e$ do:" \\

\item \hphantom{0}{\bf (Step 3: Update the target network):}
\item \hphantom{0}\hphantom{0}Copy the prediction network coefficients to the
\item[]\hphantom{0}\hphantom{0}\hphantom{0}target network.
\item Until the end of learning.
\end{enumerate}
\end{breakablealgorithm}

For comparison, the deep $Q$-learning and Sarsa learning with the {\em punishment} approach can also be obtained, denoted as ``{\em Deep Q-learning with punishment}" and ``{\em Deep Sarsa with punishment}", respectively. The {\em Deep Q-learning with punishment} and {\em Deep Sarsa with punishment} algorithms can be similarly obtained as above, except the rewards are given by \eqref{eq:rstpu} and cost function for the prediction network update is based on \eqref{eq:QcosPre}.

\section{Simulation And Discussion}\label{sec:sim}

The proposed four algorithms in Section \ref{sec:AssiDRL} are verified with simulations in this section.
For comparison, the results for the benchmark {\em max-link} buffer-aided relay selection are also shown in the simulations.

\subsection{Simulation setup}

Unless otherwise stated, the system model parameters for the buffer-aided relay network are set as following: the number of relays $K = 10$, the buffer size $L = 10$, the transmit-power-to-noise ratio $P/{\sigma}_{n}^2$ = 50 dB, the path loss exponent $\alpha = 3$. Thus the total number of states is obtained as $(4(L+1))^K$ $\approx$ 2.72$\times10^{16}$ which a very large number, making it necessary to apply the deep reinforcement learning.

The parameters related to the deep learning are set as following: the discount factor (defined in \eqref{eq:QcosPre}) $\delta=0.9$;
the exploration rate $\varepsilon$ is given by \eqref{greedy} with the initial $\varepsilon$ set to $1$, the decay factor $f = 0.999$ and $\varepsilon_{min} = 0.1$;
the learning rate of neural network is 0.01; the prediction network update iteration number $N_g = 500$; the training batch size $N_p = 32$; the target network update iteration number $N_e = 100$.

The deep learning library {\em Keras/TensorFlow} is used to build the deep neural networks. The computer with the GPU GTX-1070 is used to run the simulations.

\subsection{Simulation results}

\begin{figure}[t!]
  \centering
  \centerline{\includegraphics[scale=0.6]{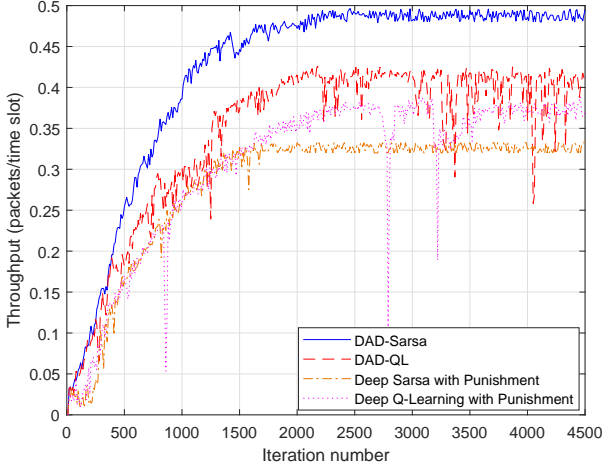}}
 \caption{ Throughput {\em vs.} training iterations, where the relay number $K=10$, the target data rate $\eta=8$ bps/Hz and the target delay $\Delta_o=6$.} \label{fig:trainfig}
\end{figure}

Fig. \ref{fig:trainfig} compares the {\em Throughput vs. Iteration} learning curves for the proposed four algorithms, where the target data rate $\eta=8$ bps/Hz, the target delay $\Delta_o=6$ and all channels are i.i.d. such that distance from all of the 10 relays to both source and destination are $5$ m. We have the following observations: \\

\begin{itemize}
    \item All of the four algorithms are able to converge. This verifies the proposed deep reinforcement learning in the delay-constrained buffer-aided relay selection.
    \item Both the decision-assisted $Q$-learning and Sarsa algorithms perform significantly better (i.e. converge to higher throughput) than their {\em punishment} counterparts. This well matches our analysis in Section \ref{sec:AssiDRL} that the punishment can be too harsh to make the learning converge to local optimums.
    \item In the decision-assisted approaches, the Sarsa learning performs better than the $Q$-learning. Similar observations are also made in Fig. \ref{fig:generalDelayfig},  \ref{fig:balanceDelayfig}, \ref{fig:generalTargetfig} and \ref{fig:balanceTargetfig}. While in the {\em punishment} approaches, we observe the opposite. This also matches our expectation because the Sarsa {\em exploits} more but {\em explores} less than its $Q$-learning counterpart. In the {\em punishment} approach, it is more important to have {\em exploration} than {\em exploitation} to avoid local optimums. While in the {\em decision-assisted} approaches, because the learning has been successfully well {\em narrowed} by giving zero target $Q$-values to invalid actions, the {\em exploitation} becomes more important.
    \item The decision-assisted Sarsa algorithm can achieve  throughput at nearly 0.5 (packet/time slot) even with delay constraint $\Delta_0=6$. This is the highest possible throughput in the two-hop relay network, making it a very attractive scheme.
\end{itemize}

\begin{figure}[t!]
  \centering
  \centerline{\includegraphics[scale=0.6]{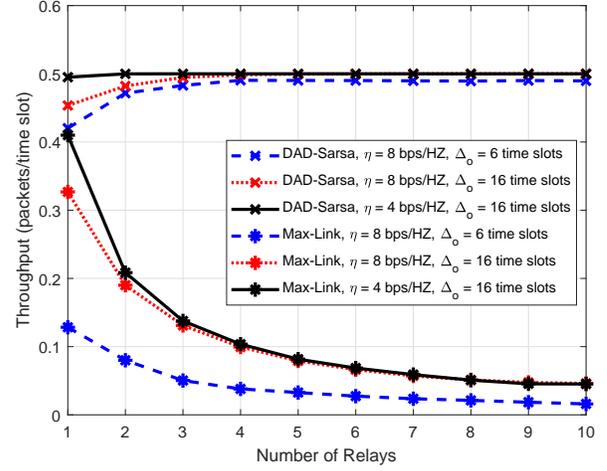}}
 \caption{ Throughput {\em vs.} Relays number comparison for the {\em DAD-Sarsa} and {\em max-link}.} \label{fig:relayNumber}
\end{figure}

Fig. \ref{fig:relayNumber} compares the throughput with respect to the relay numbers for the decision-assisted Deep Sarsa (i.e. {\em DAD-Sarsa}) and the benchmark {\em max-link} scheme. As is specified in Fig. \ref{fig:relayNumber}, several target data rates $\eta$ and target delays $\Delta_o$ are used. All other parameters are the same as those in Fig. \ref{fig:trainfig}. It is clearly shown that in all cases, the {\em DAD-Sarsa} performs significantly better than the {\em max-link}. With more relays, the {\em DAD-Sarsa} can achieve throughput close to 0.5 in all cases. On the other hand, the throughput for the {\em max-link} deteriorates dramatically with more relays. This is because the delay in the {\em max-link} increases linearly with the relay number.

\begin{figure}[t!]
  \centering
  \centerline{\includegraphics[scale=0.6]{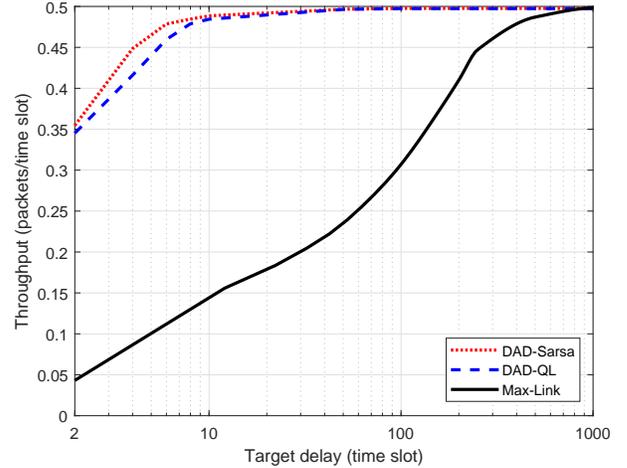}}
 \caption{  Throughput vs. Target Delay ($\Delta_o$) for i.n.i.d. channels, where the relay number $K=10$ and the target data rate $\eta=8$ bps/Hz.} \label{fig:generalDelayfig}
\end{figure}

\begin{figure}[t!]
  \centering
  \centerline{\includegraphics[scale=0.6]{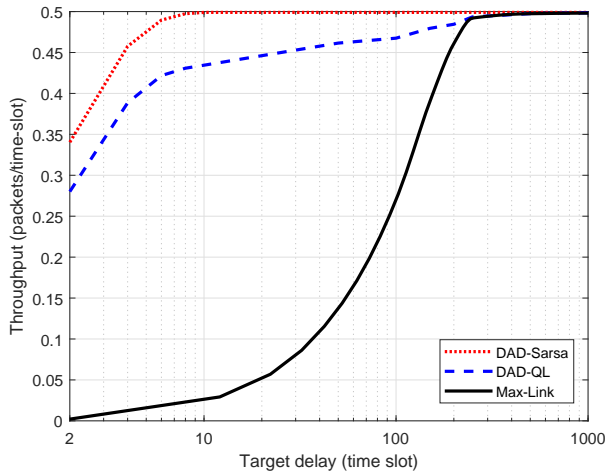}}
 \caption{  Throughput vs. Target Delay ($\Delta_o$) for i.i.d. channels, where the relay number $K=10$ and the target data rate $\eta=8$ bps/Hz.} \label{fig:balanceDelayfig}
\end{figure}

Fig. \ref{fig:generalDelayfig} shows the throughput with respect to the target delay ($\Delta_o$) for the proposed {\em DAD-Sarsa}, {\em DAD-QL} and the benchmark {\em max-link} scheme, where the target data rate $\eta=8$ bps/Hz. The independent-non-identical-distributed (i.n.i.d.) channels are considered: the locations of the source and destination nodes in the 2-dimensional map are (0, 0) and (10, 0) respectively, and the locations for the $10$ relays are (4.0, -2.6), (2.9, 2.1), (6.3, 2.5), (3.6, -1.2), (4.5, 2.1), (7.8, 0.2), (4.1, 3.5), (6.7, -2.9), (5.2, 1.8) and (7.6, 2.1), respectively. It is clearly shown that the proposed decision-assisted deep learning algorithms perform significantly better than the {\em max-link}. In the {\em max-link}, even with $\Delta_o=100$, the throughput is only about $0.3$. While for the decision-assisted learning, the throughput is already $0.35$ when $\Delta_o=2$, and quickly increases to nearly $0.5$ when $\Delta_o=10$. This clearly indicates that the {\em max-link} can only be used in the data transmission without delay constraints, but the proposed decision-assisted deep learning performs well with moderate delay constraints.

Fig. \ref{fig:balanceDelayfig} is similar to Fig. \ref{fig:generalDelayfig} except that the i.i.d. channels are considered. The comparisons in Figs. \ref{fig:generalDelayfig} and \ref{fig:balanceDelayfig} are similar only that the superiority of the {\em DAD-Sarsa} compared with the {\em DAD-QL} is more obvious in the i.i.d than in the i.n.i.d. channels.

\begin{figure}[t!]
  \centering
  \centerline{\includegraphics[scale=0.6]{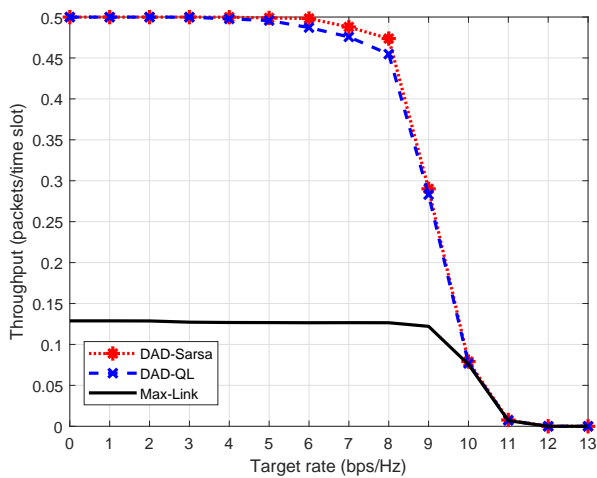}}
 \caption{ Throughput vs. Target Data Rate ($\eta$) for i.n.i.d. channels, where the relay number $K=10$ and  the target delay $\Delta_o=6$.} \label{fig:generalTargetfig}
\end{figure}

\begin{figure}[t!]
  \centering
  \centerline{\includegraphics[scale=0.6]{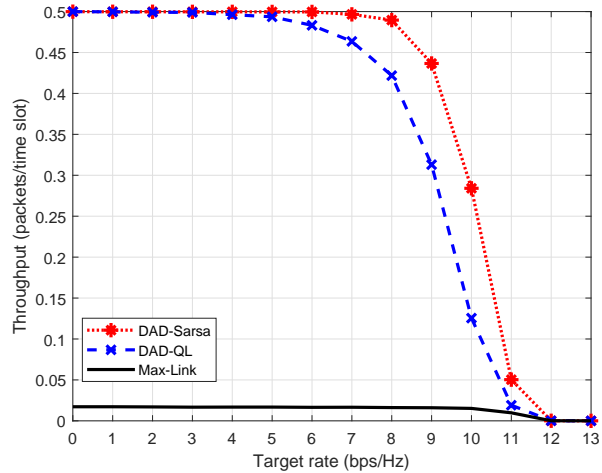}}
 \caption{ Throughput vs. Target Data Rate ($\eta$) for i.i.d. channels, where the relay number $K=10$ and  the target delay $\Delta_o=6$.} \label{fig:balanceTargetfig}
\end{figure}

Fig. \ref{fig:generalTargetfig} and \ref{fig:balanceTargetfig} compare the throughput with respect to the target data rate ($\eta$) among the {\em DAD-Sarsa}, {\em DAD-QL} and {\em max-link} schemes, for the i.n.i.d. and i.i.d. channels, respectively. The target delay $\Delta_o=6$ and all other parameters are set as same as those in Figs. \ref{fig:generalDelayfig} and \ref{fig:balanceDelayfig} correspondingly. In both figures, the proposed decision-assisted deep learning algorithms perform significantly better than the {\em max-link} scheme.

\section{Conclusion}\label{sec:con}

In this paper, we applied the deep reinforcement learning for the relay selection in the delay-constrained buffer-aided relay networks. Both the deep $Q$-learning and Sarsa were studied. Moreover, in order to explore the a-priori information from invalid actions, we investigated two methods, the {\em punishment} and {\em decision-assisted}, respectively, which can be used in either deep $Q$-learning or Sarsa, resulting in four algorithms proposed in Section \ref{sec:algorithm}.

Furthermore, because the {\em publishment} approach may lead to local optimums, it is outperformed by the {\em decision-assisted} approach. On the other hand, when the decision-assisted approach is applied, the Sarsa learning usually performs better than its $Q$-learning counterpart, though it may not be significant in some cases. With these observations, we believe the decision-assisted deep Sarsa learning is the most suitable algorithm for the relay selection in buffer-aided relay networks. This is well verified in simulations.

Finally, we highlighted that the proposed deep reinforcement algorithms can also be applied in other applications such as the routing selection, antenna selection and D2D pairing. This will be left for future study.



\balance
\bibliographystyle{ieeetr}
\bibliography{ref}

\end{document}